\documentclass[preprint,a4paper]{article}
\usepackage{latexsym}
\usepackage{amssymb}

\newtheorem {problem} {\bf Problem}
\newtheorem {theorem} {\bf Theorem}
\newtheorem {algorithm} {\bf Algorithm}

 \title{On an Algorithm for Obtaining All Binary Matrices of Special Class Related to V. E. Tarakanov's Formula}
 \author{Krasimir Yordzhev\thanks{South-West University ''N. Rilsky'', Blagoevgrad, Bulgaria, e-mail: yordzhev@swu.bg}}
 \date{}
 \begin{document}
 \maketitle
\begin{abstract}
An algorithm for obtaining all $n\times n$  binary matrices having exactly 2 units in every row and every column is described in the paper. After analysing the work of the algorithm a formula for calculating the number of these matrices has been obtained. This formula is known and has been obtained using other methods, which by their nature are purely analytical and not constructive. Thus a new, constructive proof of this known formula has been obtained.
\end{abstract}

\section{Introduction}
Everywhere in the current work $[n]$  will denote the set  $[n] =\{1,2,\ldots ,n\}$. If $M$  is a finite set, as usual $|M|$  will denote the cardinality of  $M$, and with ${\cal P}(M)$  we will be denoting the set of all permutations of the elements from  $M$. With $S_n$  we will be denoting the symmetric group,  $S_n \cong {\cal P}([n] )$. If  $\rho \in S_n$, then $\rho (i)$  is the image of $i\in [n]$  by the mapping $\rho$.

\emph{Binary} is called a matrix whose elements belong to the set $\{ 0,1\}$.

$\Lambda_n^k$-matrices we will call all $n\times n$ binary matrices in which in every row and every column there exist exactly $k$  identities and $n-k$ zeros. In the current work we will examine the following

\begin{problem}\label{zad1}
Obtain all  $\Lambda_n^2$-matrices
\end{problem}

Let us denote with $\lambda_n$  the number of all  $\Lambda_n^2$-matrices.
In  \cite{6}, the next formula is shown:
\begin{equation}\label{1}
\lambda_n = \sum_{2x_2 +3x_3 + \cdots +nx_n =n} \frac{(n!)^2}{\displaystyle \prod_{k=2}^n x_k !(2k)^{x_k}}
\end{equation}

Let us notice that formula (\ref{1}) does not give a solution to problem \ref{zad1}, it only calculates the number of the objects which are being sought. We will give a solution to problem \ref{zad1} with the help of algorithm \ref{alg1tarak} described below. Furthermore, formula  \ref{1} has been obtained by V. E. Tarakanov when examining an equivalence relation in the set of all ordered pairs of derangements  $(\rho ,\sigma )\in S_n \times S_n$. Let us remind that  $\rho ,\sigma \in S_n$  is called \emph{derangement}, if  $\rho (i) \ne \sigma (i)$  for every  $i=1,2,\ldots ,n$). In the current work we will suggest a new constructive proof of formula \ref{1} by suggesting an algorithm for obtaining all  $\Lambda_n^2$-matrices.

\section{The main results}
A \emph{partition} of the set $M$  we will be calling every family of sets  $\{ M_i \}_{i\in I}$, such that $M_i \subseteq M$ for every $i\in I$,  $ \bigcup_{i\in I} M_i =M,$ $M_i \cap M_j =\emptyset$ when $i\ne j$. The subsets  $M_i$   are called \emph{blocks} of partition. We say that a partition is of type  $1^{x_1} 2^{x_2} \ldots n^{x_n}$, if there exist exactly $x_k$  blocks with cardinality $k$, $(k=1,2,\ldots n)$. As it is well known \cite{aigner} there exists an one-to-one correspondence between the set of all types of partitions of  $M$, $|M|=n$   and the set of all solutions to the equation  $1\cdot x_1 +2\cdot x_2 +\cdots +n\cdot x_n =n$.

\begin{algorithm} \label{alg1tarak}
Obtaining all  $\Lambda_n^2$-matrices.
\end{algorithm}

1. \textbf{For} every solution to the equation  $2x_2 +3x_3 +\cdots +nx_n =n$ {\bf Do begin}

2. \textbf{For} every partition  {\bf C} of $[n]$ of type  $1^0 2^{x_2} 3^{x_3} \ldots n^{x_n}$ {\bf Do begin}

3. \textbf{For} every partition  {\bf A} of $[n]$ of type  $1^0 2^{x_2} 3^{x_3} \ldots n^{x_n}$ {\bf Do begin}

4. \textbf{For} every   $k=2,3,\ldots n$, for which  $x_k \ne 0$ {\bf Do begin}

5. Form the sets
$${\cal C} :=\{ C_1^{(k)} ,C_2^{(k)}, \ldots ,C_{x_k}^{(k)} \}$$
and
$${\cal A} :=\{ A_1^{(k)} ,A_2^{(k)}, \ldots ,A_{x_k}^{(k)} \}$$
where $C_i^{(k)} \in {\bf C} ,$ $|C_i^{(k)} |=k,$ $A_i^{(k)} \in {\bf A} ,$ $|A_i^{(k)} |=k,$ $i=1,2,\ldots ,x_k ;$

6. \textbf{For} every $\rho \in S_{x_k}$ {\bf Do begin}

7. \textbf{For} every $i=1,2,\ldots , x_k$ {\bf Do begin}

8. $l:=0;$ $L_i^{(k)} :=\emptyset$\\
/* \quad $l $ - integer; $L_i^{(k)}$ - set of integers  \quad */

9. $c_1 :=\min \{ c\; |\; c\in C_i^{(k)} \} ;$

10. \textbf{For} every permutation  $\displaystyle c_2  c_3 \ldots c_k \in {\cal P} \left( C_i^{(k)} \setminus \{ c_1 \} \right) $ {\bf Do begin}

11. \textbf{For} every permutation  $\displaystyle a_1  a_2 \ldots a_k \in {\cal P} \left( A_{\rho (i)}^{(k)} \right) $ {\bf Do begin}

12. {\bf If} $a_1 >a_2$ {\bf Then}  return to 11;

13. $l:=l+1;$ $L_i^{(k)} :=L_i^{(k)} \cup \{ l \} ;$

14. Obtain the set of three-dimensional vectors:
$$N_i^{(k)} [l]:=\left\{ \left( \begin{array}{c} c_1 \\a_1 \\a_2 \end{array} \right) ,\left( \begin{array}{c} c_2 \\ a_2 \\a_3 \end{array} \right) ,\ldots ,\left( \begin{array}{c} c_{k-1} \\a_{k-1} \\a_k \end{array} \right) ,\left( \begin{array}{c} c_k \\a_k \\a_1 \end{array} \right) \right\} ;$$

15. {\bf End } 11

16. {\bf End } 10

17. {\bf End } 7

18. Form the set  $Q_{x_k} :=L_1^{(k)} \times L_2^{(k)} \times \cdots \times L_{x_k}^{(k)} ;$

19. \textbf{For} every  $x_k$-tuple  $\overline{q} =(l_1 ,l_2 ,..., l_{x_k} )\in Q_{x_k}$ {\bf Do begin}

20. Form the set of three-dimensional vectors:
$$M[\rho ;\overline{q} ]:= \bigcup_{i=1}^{x_k} N_i^{(k)} [l_i]$$

21.  {\bf End } 19

22. {\bf End } 6

23. {\bf End } 4

24.  \textbf{For} every $(n-1)$-tuple $(\rho_2 ,\rho_3 ,\ldots \rho_n )\in S_{x_2} \times S_{x_3} \times \cdots \times S_{x_n} $ {\bf Do begin}\\
/* \quad If  $x_k =0$, then by definition $S_{x_k} =\emptyset$  and   $|S_{x_k} |=0!=1 $ \quad */

25. \textbf{For} every $(n-1)$-tuple $(\overline{q_2 }, \overline{q_3 } ,\ldots ,\overline{q_n } )\in Q_{x_2} \times Q_{x_3} \times \cdots \times Q_{x_n} $ {\bf Do begin}\\
/* \quad If $x_k =0$, then by definition  $Q_{x_k} =\emptyset$ è $|Q_{x_k} |=0!=1 $ */

26. Form the set of three-dimensional vectors:
$$R= \bigcup_{k=2}^n M[\rho_k ;\overline{q_k } ] \; ;$$\\
/*\quad If $x_k =0$, then by definition $M[\rho_k ;q_k ]=\emptyset $ è $|M[\rho_k ;q_k ]|=0!=1$ \quad */

27. From $R$ obtain only  $\Lambda_n^2$-matrix $\alpha$, such that if  $\displaystyle \left( \begin{array}{c} c\\ u\\ v \end{array} \right) \in R$, then in $\alpha$ in $c$-th column the only two units stand in rows with numbers $u$ and $v$ ;

28. {\bf End } 25

29. {\bf End } 24

30. {\bf End } 3

31. {\bf End } 2

32. {\bf End } 1

\begin{theorem} \label{th1tarak}
 With the help of algorithm \ref{alg1tarak} all  $\Lambda_n^2$-matrices are obtained, moreover during every execution of statement 27 a different  $\Lambda_n^2$-matrix is obtained.
\end{theorem}

Proof. Let  $\displaystyle w' =\left( \begin{array}{c} c' \\ u' \\ v' \end{array} \right)$ and  $\displaystyle w'' =\left( \begin{array}{c} c'' \\ u'' \\ v'' \end{array} \right)$   be two vectors from set  $R$, which is obtained during execution of statement 26 from algorithm \ref{alg1tarak}. If there exist   $i,k$ and $l$, such that $w',w'' \in N_i^{(k)} [l]$   (i.e. $w'$ and $w''$ are obtained during single execution of statement 14), then according to statements 10 and 14  $c' \ne c''$. According to statements 20 and 26 it is not possible to exist  $i,k_1 ,k_2 ,l,t$, $l\ne t$, such that  $w' \in N_i^{(k_1 )} [l],$ $w''\in N_i^{(k_2 )} [t]$   and  $w'$ and $w''$   to be from one and the same set  $R$, which is obtained during execution of statement 26. If $w'\in N_i^{(k_1 )} [l],$ $w''\in N_j^{(k_2 )} [t]$ when $i\ne j$, then   $c'$ and $c''$    are from different blocks of the partition {\bf C}, i.e. again   $c'\ne c''$. Furthermore, obviously for every $c\in [n]$  there exists a block  $C_i^{(k)} \in {\bf C}$, such that $c\in C_i^{(k)}$  and according to statements 10 and 14 there exist a natural number  $l$ and  $u,v\in [n]$, such that  $\displaystyle \left( \begin{array}{c} c\\ u\\ v \end{array} \right) \in N_i^{(k)} [l]$. Therefore the first elements of every vector in the set  $R$ during single execution of statement 26 correctly indicate a number of a column in a matrix.

Let  $a\in [n]$. Since statements 11,12,14 and 26 are inner for loops with numbers 2 and 3, then according to these statements there exist unique three-dimensional vectors   $w'$ and $w''$, $w'\ne w''$  from the set $R$  during single execution of statement 26, such that $a$  be at second position in $w'$  and at third position in  $w''$. Therefore in the matrices which are obtained in statement 27 there are exactly two units in every row and according to statement 27 exactly two units in every column.

We assume that there exist two sets   $R_1$ and $R_2$, obtained during different executions of statement 26, which are equal to each other. Then   $R_1 =R_2  =N_1^{(2)} [l_1 ]\cup \cdots \cup N_{x_2}^{(2)} [l_{x_2} ] \cup \cdots \cup N_{x_n}^{(n)} [l_{x_n} ] $, which is impossible, bearing in mind the courses of action of statements 14, 20 and 26, and with what objects they work during each of their executions. Therefore during every execution of statement 27 a different   $\Lambda_n^2$-matrix is obtained.

Let us denote with $T$  the set of the matrices obtained during the work of statement 27. Since in  $T$ there are not any duplicate elements, then  $T\subseteq \Lambda_n^2$.

Let $\alpha $  be an arbitrary  $\Lambda_n^2$-matrix. We will show that   $\alpha \in T$. Let in the beginning $\alpha$  have not any marked rows and columns. In $\alpha$  we build a \emph{walk}, starting from the upper ''1'' of the leftmost unmarked column of the matrix, we continue to the next below ''1'' in the column, after that we reach the second ''1'' from the same row, then the second ''1'' from the same column and so on until we return to the initial cell. Moving on this walk we mark the rows and the columns we pass, and when passing through  $c$-th column we obtain the vector $\displaystyle \left( \begin{array}{c} c\\ u\\ v \end{array} \right) $  (we have first passed through the ''1'' in   $u$-th row, then through the ''1'' in  $v$-th row). We continue the same process with the unmarked rows and columns until we mark the whole matrix. Let $c_1$  be the minimum number among the numbers standing at first position of the vectors corresponding to a particular walk. Since the beginning of the loop starts from the upper ''1'' of the leftmost column of the walk, then for the vector $\displaystyle \left( \begin{array}{c} c_1 \\ u\\ v \end{array} \right) $ the inequality $u<v$ is satisfied. Therefore for every walk we obtain a set of vectors, which are obtained in statement 14 as well. Uniting the sets of all vectors which are obtained from walks in $\alpha$  with the same length, we obtain a set, which is obtained in statement 20 as well. The union of all such sets corresponding to a particular matrix, is a set, which is obtained in statement 26 as well. Therefore  $\left| \Lambda_n^2 \right| \le \left| T\right|$, whence the assertion of the theorem follows.

\hfill $\Box$

With the help of Algorithm \ref{alg1tarak} and Theorem \ref{th1tarak} we will also prove the next theorem, proven in \cite{6} in a different way.
\begin{theorem} \label{th2tarak} \cite{6}
The number $\lambda_n$ of all $\Lambda_n^2$-matrices  is given by the formula:

$$\lambda_n = \sum_{2x_2 +3x_3 + \cdots +nx_n =n} \frac{(n!)^2}{\displaystyle \prod_{k=2}^n x_k !(2k)^{x_k}} $$
\end{theorem}

Proof. Let us denote with  $\mu_j$  the number of the iterations of loop with number $j$  in Algorithm  \ref{alg1tarak}. Since Algorithm  \ref{alg1tarak} has been maid so that every time during execution of statements 14, 20, 26 and therefore also statement 27 (it follows from Theorem \ref{alg1tarak}) different objects are obtained, then
\begin{equation} \label{th2f1}
\lambda_n =\mu_1 \mu_2 \mu_3 \mu_{24} \mu_{25} ,
\end{equation}

Is easy to see that
$$\displaystyle \mu_1 =\sum_{2x_2 +3x_3 + \cdots +nx_n =n} 1 .$$

According to a well-known formula \cite[(3.15.ii)]{aigner} the next equation is true:
$$\displaystyle \mu_2 =\mu_3 =\frac{n!}{\displaystyle \prod_{k=2}^n x_k ! (k!)^{x_k} }$$

For $\mu_{24}$  and $\mu_{25}$  we obtain respectively:
$$\displaystyle \mu_{24} =\prod_{k=2}^n \left| S_{x_k} \right| =\prod_{k=2}^n x_k !$$
$$\displaystyle \mu_{25} =\prod_{k=2}^n |Q_{x_k} |=\prod_{k=2}^n \left( \prod_{j=1}^{x_k} \left| L_j^{(k)} \right| \right) =\prod_{k=2}^n \left( \prod_{j=1}^{x_k} \mu_{10} \mu_{11} \right) =$$
$$\displaystyle =\prod_{k=2}^n \left(\prod_{j=1}^{x_k} (k-1)! \cdot \frac{k!}{2} \right) =\prod_{k=2}^n \frac{(k!)^{2x_k } }{(2k)^{x_k} } $$

Substituting in  (\ref{th2f1}) we obtain the assertion of the theorem as well.

\hfill $\Box$
\section{Conclusion}
Comparing the proof of theorem  \ref{th2tarak} to formula (\ref{1}) we are convinced of the high efficiency  of algorithm \ref{alg1tarak}. This is practically a proof of the fact that algorithm \ref{alg1tarak} works exactly as much as necessary.

\end{document}